\documentclass[12pt]{article}
\usepackage[utf8]{inputenc}
\usepackage{amsmath,amsthm,amssymb,amsfonts}
\usepackage{prettyref}
\usepackage{fancyhdr}
\usepackage[a4paper]{anysize}
\usepackage[light]{draftcopy}
\usepackage{comment}
\usepackage{tikz,pgf}
\usepackage{pgfplotstable}
\usepackage{filecontents}

\usepackage[font=small,labelfont=bf,tableposition=top]{caption}

\DeclareCaptionLabelFormat{andtable}{#1~#2  \&  \tablename~\thetable}

\renewcommand{\Im}{\text{Im}}
\newif\ifdebug

\ifdebug

\else

\fi

\theoremstyle{plain}

\newtheorem{thm}{Theorem}

\theoremstyle{definition}

\newrefformat{eq}{(\ref{#1})}
\newrefformat{tab}{Table \ref{#1}}

\author{Nugzar Shavlakadze\footnote{Tbilisi State University
    A. Razmadze Mathematical Institute, Tamarashvili str. 6, 0177,
    Tbilisi, Georgia e-mail:nusha@rmi.ge, nusha1961@yahoo.com}\and
  Nana Odishelidze\footnote{Department of Computer Sciences
    Faculty of Exact and Natural Sciences,Iv.Javakhishvili Tbilisi
    State University, 2, University st., 0143,Georgia.  Tfn:
    99532304784. E-mail: nana\_georgiana@yahoo.com} \and
  Francisco Criado-Aldeanueva\footnote{Department of Applied Physics
    II, Polytechnic School, Malaga University, Campus Teatinos, s/n
    (29071), Spain. Tfn: +34952132849. E-mail:
    fcaldeanueva@ctima.uma.es}}

\title{Exact solutions of some singular integro-differential equations
  related to adhesive contact problems of elasticity theory}

\date{}

\begin{document}

\maketitle

\begin{abstract}
  The problem of constructing an exact solution of singular
  integro-differential equations related to problems of adhesive
  interaction between elastic thin semi-infinite homogeneous patch and
  elastic plate is investigated. For the patch loaded with horizontal
  forces the usual model of the uniaxial stress state is valid. Using
  the methods of the theory of analytic functions and integral
  transformation the singular integro-differential equation is reduced
  to the Riemann boundary value problem of the theory of analytic
  functions.  The exact solution of this problem and asymptotic
  estimates of tangential contact stresses are obtained.\bigskip

  \smallskip

\end{abstract}

\section{Introduction}

Exact or approximate solutions of static contact problems for
different domains, reinforced by elastic thin inclusions and patches
of variable stiffness were obtained and the behavior of the contact
stresses at the ends of the contact line has been investigated as a
function of the geometrical and physical parameters of these elements
\cite{1,7,8,6,4,3,12,5,2,13,14,18,15,16,17,10,11,9}.  One model
assumes continuous interaction, while the other assumes the adhesive
contact of thin-shared elements (stringers or inclusions) with massive
deformable bodies. In \cite{19}, a finite-length stringer is attached
to a thin elastic sheet subjected to plane stress. The two different
materials are joined along the entire stringer length by a thin
uniform elastic adhesive layer assumed to be in pure shear state. The
bending is neglected and the interaction between the sheet and the
stringer is idealized as a line loading of the sheet. In \cite{20}, an
elastic semi-infinite plate is strengthened by an elastic finite
stringer. The contact between the plate and the stringer is achieved
by a thin glue layer. Asymptotic estimates, exact and approximate
solutions of the associated integro-differential equation are
obtained.

In the present paper the exact solution of singular
integro-differential equations related to the problems of adhesive
interaction between an elastic thin semi-infinite homogeneous patch
and an elastic plate is obtained. From the physical point of view,
consideration of the case of the adhesive interaction between the
plate and the stringer is interesting since the boundedness of the
unknown tangential contact stress near the ends of the stringer is
proved by rigorous mathematical methods. The limit transition (case A)
corresponds to the case of rigid contact, in which the obtained
solution, i.e. tangential contact stress has a singularity at the end
of the stringer.

\section{Formulation of the problem and reduction to the
  integro-differential equation}

Let a semi-infinite patch with modulus of elasticity $E_1(x)$,
thickness $h_1(x)$ and Poisson's coefficient $\nu_1$ be attached to
the plate ($E_2,\nu_2$), which is in the condition of a plane
deformation. It is assumed that the horizontal stresses with intensity
$\tau_0(x)$ act on the patch along the ox-axis. In the horizontal
direction the patch is compressed or stretched like a rod being in
uniaxial stress state. The contact between the plate and patch is
achieved by a thin glue layer with width $h_0$ and Lame's constants
$\lambda_0$, $\mu_0$.

The adhesive contact condition has the form \cite{19}
\begin{equation}
  u_1(x) - u_2(x,0) = k_0\tau( x),\qquad x>0\label{eq:1.1}
\end{equation}
where $u_2(x,y)$ and $u_1(x)$ are displacements of the plate points
and displacements of the patch points, respectively. $\tau(x)$ is
unknown tangential contact stresses and $k_0 = h_0/\mu_0$.

We have to define the law of distribution of tangential contact
stresses $\tau(x)$ on the line of contact, the asymptotic behavior of
these stresses at the end of the patches.

According to the equilibrium equation of patch elements and Hooke's
law we have:
\begin{equation}
  \frac{du_1(x)}{dx}
  = \frac{1}{E(x)}\int_0^x [\tau(t) - \tau_0 (t)]dt,
  \qquad
  x > 0,
  \label{eq:1.2}
\end{equation}
and the equilibrium equation of the patch has the form
\begin{equation}
  \begin{gathered}
    \int_0^\infty [\tau(t) -\tau_0(t)]dt = 0,\\
    E(x) = \frac{E_1(x)h_1(x)}{1 -\nu_1^2}
  \end{gathered}
  \label{eq:1.3}
\end{equation}

According to known results for plate \cite{21}, the horizontal
deformation of the points of the $OX$ axis have the form
\begin{equation}
  \frac{du_2(x,0)}{dx}
  = \frac{b}{\pi}\int_0^\infty\frac{\tau(t)dt}{t-x}
  \label{eq:1.4}
\end{equation}
where
\begin{equation*}
  b= \frac{2(1 - \nu_2^2 )}{E_2}
\end{equation*}

Introducing the notation
\begin{equation*}
  g(x) = \int_0^x [\tau(t) - \tau_0(t)]dt,
\end{equation*}
from (\ref{eq:1.1}), (\ref{eq:1.2}) and (\ref{eq:1.4}) we obtain the
following integro-differential equation
\begin{equation}
  \frac{g(x)}{E(x)}
  - \frac b\pi \int_0^\infty\frac{g'(t)\,dt}{t-x}
  - k_0 g''(x) = f(x), \qquad x > 0
  \label{eq:1.5}
\end{equation}
where
\begin{equation*}
  f(x)
  = \frac b\pi \int_0^\infty \frac{\tau_0(t)\,dt}{t-x}
  + k_0 \tau_0'(x)
\end{equation*}
with the condition
\begin{equation}
  g(\infty) = 0
  \label{eq:1.6}
\end{equation}
Thus, the above posed boundary contact problem is reduced to the
singular integro-differential equation (\ref{eq:1.5}) with the
condition (\ref{eq:1.6}).

The solutions of equation (\ref{eq:1.5}) under the condition
(\ref{eq:1.6}) can be sought in the class of functions:
$g,g'\in H[0,\infty)$, $g''\in H (0,\infty)$
\cite{22}.

We assume that the function $\tau_0(x)$ is continuous in the Holder's
sense and $\tau_0(x)$ has a first order continuous derivative.

\section{Exact solution of singular integro-differential equations}

Suppose that a plate on a semi-infinite interval is reinforced with a
homogeneous patch and is free of external loads. The contact between
the plate and the patch is carried out through a thin layer of glue.

The problem means determination of contact stresses when a horizontal
force $T$ applies at one end of the patch (at a point $x = 0$).
$E(x) = E_0 = \text{const}$.

The equation (\ref{eq:1.5}) and the boundary conditions (\ref{eq:1.6})
take the form
\begin{gather}
  \varphi(x)
  - \frac\lambda\pi \int_0^\infty\frac{\varphi'(t)dt}{t-x}
  - k\varphi''(x) = 0, \qquad x > 0
  \label{eq:2.1}\\
  \varphi(0) = T,\quad \varphi(\infty) = 0,
  \label{eq:2.2}
\end{gather}
where
\begin{equation*}
  \varphi(x) = T - \int_0^x\tau(t)dt,\quad
  \lambda = b E_0,\quad
  k = k_0 E_0.
\end{equation*}
The solution of equation (\ref{eq:2.1}) is sought in the class of
functions $\varphi , \varphi'\in H[0,\infty)$,
$\varphi''\in H(0,\infty)$.

By a generalized Fourier transform with the convolution theorem
\cite{23}, from (\ref{eq:2.1}), (\ref{eq:2.2}) we arrive at a Riemann
problem
\begin{equation}
  \Phi^+(s)\left[
    1 + \lambda|s| + ks^2\right]
  = F^-(s) - i\lambda T \text{sgn} s - k\varphi'(0) + i k s T,
  \label{eq:2.3}
\end{equation}
where $\Phi^+(s)$ and $F^-(s)$ are Fourier transforms of functions
\begin{equation*}
  \varphi_0(x) = \begin{cases}
    \varphi(x), & x\ge 0\\
    0, & x<0
  \end{cases}
  \qquad
  \text{ and }
  \qquad
  f(x) = \begin{cases}
    0, & x\ge 0\\
    \frac{\lambda}{\pi}\int_{-\infty}^\infty\frac{\varphi_0'(t)dt}{t-x}
    - k\varphi_0''(x), & x<0
  \end{cases}
\end{equation*}

\section{Case A}

a) For $k \ge  0$, the coefficient of the problem (\ref{eq:2.3}) can be represented in the form
\begin{equation*}
  1 + \lambda|s| + ks^2
  = \frac{1 + \lambda |s| + ks^2}{
    \sqrt{1 + \lambda^2 s^2}
    \sqrt{1 + \widetilde{k}^2 s^2}
  }
  \sqrt{\lambda s + i}
  \sqrt{\lambda s -i}
  \sqrt{\widetilde{k} s + i}
  \sqrt{\widetilde{k} s - i}, \quad
  \widetilde{k} =\frac{k}{\lambda}
\end{equation*}
and we consider the canonical solution of the problem of linear
conjugation
\begin{equation}
  X^+(s) =\frac{
    1 + \lambda |s| + ks^2
  }{
    \sqrt{1 + \lambda^2 s^2}
    \sqrt{1 + \widetilde{k}^2 s^2}
  }
  X^-(s)
  \label{eq:2.4}
\end{equation}

Everywhere, we mean by functions of type $\sqrt{\lambda z +i}$ and
$\sqrt{\lambda z - i}$ the branches that are analytic in planes with
cuts along the rays, drawn from the points $z=-i/\lambda$ and
$z=i/\lambda$ respectively, in the $OX$ direction, and which take
positive and negative values respectively on the upper side of the
cut. With this choice of branches the function $\sqrt{1+\lambda^2
  z^2}$ is analytic in the strip $-1/\lambda < \Im z < 1/\lambda$ and
takes a positive value on the real axis.

The function
\begin{equation*}
  X(z) = \exp\left\{
    \frac{1}{2\pi i}
    \int_{-\infty}^\infty
    \ln\frac{1 + \lambda |s| + k s^2}{
      \sqrt{1 + \lambda^2 s^2}
      \sqrt{1 + \widetilde{k}^2 s^2}
    }
    \frac{ds}{s-z}
  \right\}
\end{equation*}
satisfies the boundary condition (\ref{eq:2.4}), does not vanish
anywhere and $X^\pm(\infty) = 1$.

Representing the boundary condition (\ref{eq:2.3}) in the form
\begin{multline*}
  \Phi^+(s)\sqrt{(\lambda s + i)(\widetilde{k}s + i)} X^+(s)\\
  = \frac{F^-(s)X^-(s)}{\sqrt{(\lambda s -i)(\widetilde{k}s - i)}}
  +\frac{
    (-i\lambda T \text{sgn} s - k\varphi'(0) + ikTs) X^-(s)
  }{  
    \sqrt{(\lambda s -i)(\widetilde{k}s - i)}
  }
\end{multline*}
we get
\begin{multline*}
  \Phi(z)\sqrt{(\lambda z + i)(\widetilde{k}z + i)} X(z)\\
  = - \frac{\lambda T}{2\pi}
  \int_{-\infty}^\infty
  \frac{X^-(s) \text{sgn} s}{\sqrt{(\lambda s - i)(\widetilde{k}s -i)}}\frac{ds}{s-z}
  - \frac{k\varphi'(0)}{2\pi i}
  \int_{-\infty}^\infty \frac{X^-(s)}{
    \sqrt{(\lambda s - i)(\widetilde{k}s -i)}
  }\frac{ds}{s - z}\\
  + \frac{kT}{2\pi} \int_{-\infty}^\infty
  \frac{s X^-(s)}{
    \sqrt{(\lambda s - i)(\widetilde{k}s - i)}
  } \frac{ds}{s - z}
\end{multline*}
and based on the well-known Cauchy formula
\begin{multline*}
  \Phi(z)
  = \frac{-\lambda T}{
    \pi X(z) \sqrt{(\lambda z + i)(\widetilde{k}z + i)}
  } \int_0^\infty
  \frac{X^-(s)}{
    \sqrt{(\lambda s - i)(\widetilde{k}s - i)}
  }\frac{ds}{s-z}\\
  + \frac{i\sqrt{k}T}{
    2X(z) \sqrt{(\lambda z + i)(\widetilde{k}z + i)}
  },\qquad
  \text{Im} z > 0
\end{multline*}
The boundary value of function $K(z)=-T-i z\Phi(z)$ is Fourier
transform of function $\varphi'(x)$.

We investigate the behavior at infinity of the following function:
\begin{multline}
  K(z) = -T
  - \frac{\lambda T z}{
    \pi iX(z)\sqrt{(\lambda z + i)(\widetilde{k} z + i)}
  } \int_0^\infty 
  \frac{X^-(s)ds}{
    \sqrt{(\lambda s - i)(\widetilde{k}s - i)}(s-z)
  }\\
  + \frac{z \sqrt{k} T}{
    2 X(z)\sqrt{(\lambda z + i)(\widetilde{k}z + i)}
  }
  \label{eq:2.5}
\end{multline}
Introducing the notations
\begin{equation*}
  \begin{aligned}
    K_1(z)
    &=\frac{\lambda Tz}{\pi i X(z)\sqrt{(\lambda z + i)(\widetilde{k}z + i)}}
    \int_0^\infty
    \frac{X^-(s)\,ds}{\sqrt{(\lambda s - i)(\widetilde{k}s - i)}(s-z)},\\
    K_2(z)
    &= \frac{z \sqrt{k} T}{2 X(z)\sqrt{(\lambda z + i)(\widetilde{k}z + i)}}
  \end{aligned}
\end{equation*}
and the change of variable $z = -1/\xi$, $s=-1/t_0$ for function
$K_1(z)$ gives
\begin{equation}
  K_1^*(\xi)
  = \frac{
    - \lambda T\xi
  }{
    \pi iX^*(\xi)\sqrt{(\lambda - i\xi)(\widetilde{k} - i\xi)}
  }
  \int_{-\infty}^0
  \frac{X^{-*}(t_0)\,dt_0}{
    \sqrt{(\lambda+i t_0)(\widetilde{k} + i t_0)}(t_0-\xi)
  },
  \label{eq:2.6}
\end{equation}
where $K_1^*(\xi) = K_1(z)$, $X^*(\xi) = X(z)$. By virtue of known
results \cite{22}, $K_1^*(\xi) = O(\xi ln \xi)$, $\xi \to 0$ and
respectively $K_1(z) = O(|z|^{-(1-\epsilon)})$, $|z| \to \infty$
($\epsilon$ arbitrarily small positive number).

Since $K_2(\infty) = T/2$, , the function
$\widetilde{K}_1^+(z) = K(z) + T/2$ is holomorphic in half-plate
$\text{Im} z > 0$ and vanishes at infinity as $|z|^{-(1-\epsilon)}$,
$|z|\to\infty$.

Consequently, unknown tangential contact stresses are determined by
the formula
\begin{equation}
  \tau(x) = \varphi' (x)
  = \frac{1}{2\pi}
  \int_{-\infty}^\infty \widetilde{K}_1(t) e^{-itx} \,dt,
  \label{eq:2.7}
\end{equation}
and it is bounded, when $x \to 0^+$.

By limiting transition $k \to 0$ from (\ref{eq:2.6}) we have
\begin{equation*}
  \begin{gathered}
    K_1^*(\xi)
    = \frac{\lambda T \sqrt{\xi}}{\pi iX^*(\xi) \sqrt{\lambda - i\xi}}
    \int_{-\infty}^0 \frac{X^{-*}(t_0)d t_0}{\sqrt{t_0 (\lambda + i t_0)} (t_0 - \xi)},\\
    K_1^*(\xi) = -T + O(\xi^{1/2-\delta}), \qquad
    0 < \delta < 1/2, \qquad
    \xi \to 0,\\
    K_1(z) + T = O(|z|^{-(1/2 +\delta)}),\qquad
    |z| \to\infty,\qquad
    K_2(z) = 0.
  \end{gathered}
\end{equation*}

The tangential contact stresses are determined by the formula
\begin{equation*}
  \tau(x) = \varphi'(x)
  =\frac{1}{2\pi} \int_{-\infty}^\infty \widetilde{K}_2(t)e^{-itx} dt
\end{equation*}
where the function $\widetilde{K}_2^+(z) = K(z) + 2T$ is holomorphic
in half-plate $\text{Im} z > 0$ and vanishes at infinity as
$|z|^{-(1/2+\delta)}$, $|z| \to\infty$.

Therefore, tangential contact stresses $\tau(x)$, when $x \to 0^+$, has a singularity less than $1/2$. This result
matches to results from \cite{24,13,15}.

\section{Case B}
b) Let $k > k_1 > 0$, then the solution of problem (\ref{eq:2.3}) can
be represented in another form. The coefficient of the problem can be
written in the form
\begin{equation*}
  1 + \lambda |s| + ks^2 =
  \frac{1 + \lambda |s| + ks^2}{1 + ks^2}
  (1 - i\sqrt{k}s )(1 + i\sqrt{k}s )
\end{equation*}
and the canonical solution of the problem of linear conjugation
\begin{equation*}
  X^+(s) =\frac{1 + \lambda |s| + ks^2}{1 + ks^2}X^-(s)
\end{equation*}
has the form
\begin{equation*}
  X(z) = \exp\left\{
    \frac{1}{2\pi i}
    \int_{-\infty}^\infty
    \ln\frac{1 + \lambda |s| + ks^2}{1 + ks^2}
    \frac{ds}{s - z}
  \right\}.
\end{equation*}
The function $X(z)$ does not vanish anywhere and $X^\pm(\infty) = 1$.

Representing the boundary condition (\ref{eq:2.3}) in the form
\begin{equation*}
  \Phi^+(s)(1 - i\sqrt{k}s) X^+(s)
  =\frac{F^-(s) X^-(s)}{1+ i\sqrt{k}s}
  -\frac{i\lambda T \text{sgn} s + k\varphi'(0) - i k T s}{\sqrt{1+ i \sqrt{k}s}}
\end{equation*}
we get
\begin{multline*}
  \Phi(z)(1 - i\sqrt{k}z ) X(z)
  = -\frac{\lambda T}{2\pi}
  \int_{-\infty}^\infty
  \frac{X^-(s) \text{sgn} s}{1 + i\sqrt{k}s}
  \frac{ds}{s - z}\\
  -\frac{k\varphi'(0)}{2\pi i}
  \int_{-\infty}^\infty
  \frac{X^-(s)}{  1 + i\sqrt{k}s}
  \frac{ds}{s - z}
  + \frac{kT}{2\pi}
  \int_{-\infty}^\infty
  \frac{sX^-(s)}{1 + i\sqrt{k}s}
  \frac{ds}{s - z}
\end{multline*}
and based on the well-known Cauchy formula
\begin{equation*}
  \Phi(z)
  = -\frac{\lambda T}{\pi X(z)(1 - i\sqrt{k}z)}
  \int_0^\infty
  \frac{X^-(s)}{1 + i\sqrt{k} s}
  \frac{ds}{s - z}
  + \frac{\sqrt{k}T}{2X(z)(1 - i\sqrt{k}z)},\qquad
  \text{Im} z > 0
\end{equation*}
The boundary value of function $K^\circ(z) = -T-iz\Phi(z)$ is Fourier
transform of function $\varphi'(x)$.

We investigate the behavior of the function
\begin{equation*}
  K^0(z) = -T
  +\frac{i\lambda T z}{\pi X(z)(1-i\sqrt{k}z)}
  \int_0^\infty
  \frac{X^-(s)ds}{(1+i\sqrt{k}s)(s-z)}
  -\frac{iz\sqrt{k}T}{2X(z)(1 - i\sqrt{k}z)}
\end{equation*}
at infinity. Introducing the notations
$K_1^0(z) =\frac{i\lambda T z}{\pi X(z)(1-i\sqrt{k}z)} \int_0^\infty
\frac{X^-(s)ds}{(1+i\sqrt{k}s)(s-z)} $,
$K_2^0(z) = \frac{- iz\sqrt{k}T}{2X(z)(1 - i\sqrt{k}z)}$, and the
change of variable $z = -1/\xi$, $s=-1/t_0$ in function $K_1^0(z)$
gives
\begin{equation*}
  K^{0*}_1(\xi) =
  \frac{\lambda T\xi}{\pi X^*(\xi)(\xi + i \sqrt{k})}
  \int_{-\infty}^0
  \frac{X^{-*}(t_0)dt_0}{(t_0 - i \sqrt{k})(\xi - t_0 )},
\end{equation*}
where $K_1^*(\xi) = K_1(z)$, $X^*(\xi) = X(z)$.

It's obvious that $K^{0*}_1(\xi) = O(\xi ln \xi )$, $\xi \to 0$ and
$K_1^0(z) = O(|z|^{- (1-\epsilon)}))$ $|z| \to\infty$,
$K_2^0(\infty )=\frac T2$ ($\epsilon$ arbitrarily small positive
number).

Consequently, unknown contact stresses are determined by the formula
\begin{equation}
  \tau(x) = \varphi' (x)
  =\frac{1}{2\pi}
  \int_{-\infty}^\infty
  \widetilde{K}_3(t) e^{-itx} dt,
  \label{eq:2.8}
\end{equation}
where the function $\widetilde{K}_3^+(z) = K^0(z) + T/2$ is
holomorphic in half-plate $\text{Im} z > 0$ and vanishes at infinity
as $|z|^{-(1-\epsilon )}$, $|z| \to\infty$. Therefore, tangential
contact stresses, defined by formula (\ref{eq:2.8}), are bounded when
$x \to 0^+$.

Thus, it is proved the following theorem
\begin{thm} Integro-differential equation
  (\ref{eq:2.1})-(\ref{eq:2.2}) has the solution, which is represented
  effectively by formulas (\ref{eq:2.7})-(\ref{eq:2.8}) and
  $\varphi'(x) = O(1)$, $x \to 0^+$.
\end{thm}

In table \ref{fig:1} and figure \ref{tab:1} the dependence of the
tangential contact stress $\tau(x)$ at the point $x=0$ with the value
of $k$ is presented for the following physical and geometric
parameters of the problem: module of elasticity $E_2=95\cdot
10^9\,\text{Pa}$ and Poisson's coefficient $\nu_2=0.3$ of semi-plate
material; module of elasticity $E_1=120\cdot 10^9\,\text{Pa}$ and
Poisson's coefficient $\nu_1=0.5$ of stringer material and thickness
of stringer $h_1=5\cdot 10^{-2}\,\text{m}$; in the table value of
number $k=h_0E_0/\mu_0$ is defined for various values of $h_0$,
$\mu_0$, for example, we obtain $k=3.42\cdot 10^{-2}\,\text{m}$ in
case of shear module $\mu_0=0.117\cdot 10^9\,\text{Pa}$ and thickness
of glue layer $h_0=5\cdot 10^{-4}\,\text{m}$.

Calculations show that a decrease of thickness or an increase of the
shear modulus of the adhesive material, i.e. a decrease of the number
$k$, corresponds to tend to the rigid contact of the stringer with the
plate, at which the tangential contact stress tends to infinity at the
endpoints of the stringer.

\begin{figure}[htbp]
  \centering

  \pgfplotsset{width=8cm,height=7cm}

  \begin{tikzpicture}      
    \begin{axis}[
        xlabel=$K$,
        ylabel=$\tau$
      ]
      \pgfplotstableread[col sep=comma]{data.csv}\loadeddata
      \addplot  table[y=tau,x=K,mark=none,color=black] {\loadeddata}; 
    \end{axis}
  \end{tikzpicture}
  \hfill
  \begin{tabular}[b]{|c|c|}\hline
    $k$ ($m$) & $\tau(0)$ ($N/Dm^2$)\\\hline
    $3.42\cdot10^{-2}$ & $1.5138$\\
    $1.52\cdot10^{-2}$ & $1.6279$\\
    $0.55\cdot10^{-2}$ & $2.5806$\\
    $0.25\cdot10^{-2}$ & $3.5236$\\
    $1.42\cdot10^{-3}$ & $6.9176$\\
    $0.55\cdot10^{-3}$ & $20.864$\\
    $0.25\cdot10^{-3}$ & $35.768$\\
    $1.0 \cdot10^{-4}$ & $42.616$\\
    $0.7 \cdot10^{-4}$ & $46.568$\\
    $1.0 \cdot10^{-5}$ & $56.545$\\
    $0.5 \cdot10^{-5}$ & $58.325$\\\hline
  \end{tabular}
  \captionlistentry[table]{A table beside a figure}\label{fig:1}
  \captionsetup{labelformat=andtable}
  \caption{Dependence of tangential contact stress $\tau(x)$ at $x = 0$ with the value of $k$}\label{tab:1}
\end{figure}

\section{Conclusion}


In this paper the well-known method of Winer-Hopf is used for solving
a Riemann problem. We made the factorizations for specific
coefficients related to the investigated integro-differential
equation, whose effective solutions and the asymptotic estimates are
obtained.

The case $k=0$ corresponds to the absolute rigid contact between the
elastic plate and patch. $k\to 0$ means that adhesive contact tends to
rigid contact between the elastic plate and patch. In case B, the
integro differential equation is considered for $k> k_1 > 0$. The
corresponding factorization of the coefficient of the Riemann problem
was carried out and the exact solution and asymptotic estimates were
obtained.

In case A, the same integro-differential equation is considered for
$k\ge 0$. Here the other factorization of the coefficient is carried
out to make a limit transition ($k\to 0$) and to compare the result
with the known results from \cite{24,13,15}.

The mechanical result of this study is the following: in condition of
the rigid contact ($k=0$) between elastic plate and patch the
tangential contact stresses at the end (in point $x=0$) of elastic
patch has a singularity less than $1/2$; the tangential contact
stresses is bounded in case of the adhesive contact among them ($k\neq
0$).


\begin{thebibliography}{10}

\bibitem{1}
V.~M. Aleksandrov and S.~M. Mkhitaryan.
\newblock {\em Contact problems for Bodies with thin coverings and layers}.
\newblock Nauka, Moscow, 1983.
\newblock in Russian.

\bibitem{7}
Yu.~A. Antipov.
\newblock Effective solution of a {P}randtl-type integro-differential equation
  on an interval and its application to contact problems for a strip.
\newblock {\em J. Appl. Math. Mech.}, 57(3):547--556, 1993.

\bibitem{8}
Yu.~A. Antipov and N.~Kh. Arutyunyan.
\newblock A contact problem for an elastic layer with cover plates in the
  presence of friction and adhesion.
\newblock {\em J. Appl. Math. Mech.}, 57(1):159–170, 1993.

\bibitem{6}
Yu.~A. Antipov and N.~G. Moiseev.
\newblock Exact solution of the two-dimensional problem for a composite plane
  with a cut that crosses the interface line.
\newblock {\em J. Appl. Math. Mech.}, 55(4):531--539, 1999.

\bibitem{4}
R.~D. Bantsuri.
\newblock A certain boundary value problem in analytic function theory.
\newblock {\em Sakharth. SSR Mecn. Akad. Moambe}, 73:549--552, 1974.

\bibitem{3}
R.~D. Bantsuri.
\newblock The contact problem for an anisotropic wedge with an elastic
  fastening.
\newblock {\em Dokl. Akad. Nauk SSSR}, 222(3):568--571, 1975.

\bibitem{12}
R.~D. Bantsuri and N.~Shavlakadze.
\newblock The contact problem for an anisotropic wedge-shaped plate with an
  elastic fastening of variable stiffness.
\newblock {\em J. Appl. Math. Mech.}, 66(4):645–650, 2002.

\bibitem{23}
F.~D. Gakhov and Yu.~I. Cherskii.
\newblock {\em Equations of convolution type}.
\newblock ``Nauka'', Moscow, 1978.

\bibitem{20}
Otar Jokhadze, Sergo Kharibegashvili, and N.~Shavlakadze.
\newblock Approximate and exact solution of a singular integro-differential
  equation related to contact problem of elasticity theory.
\newblock {\em J. Appl. Math. Mech.}, 82(1):114--124, 2018.

\bibitem{24}
A.I. Kalandiya.
\newblock {\em Mathematical Methods of Two-Dimensional Elasticity}.
\newblock Mir Publishers, Moscow, 1975.

\bibitem{19}
J.L. Lubkin and L.C. Lewis.
\newblock Adhesive shear flow for an axially-loaded finite stringer bonded to
  an infinite sheet.
\newblock {\em Quart. J. Mech. Appl. Math.}, 4(23):521--533, 1970.

\bibitem{21}
N.~I. Muskhelishvili.
\newblock {\em Some basic problems of the mathematical theory of elasticity.
  {F}undamental equations, plane theory of elasticity, torsion and bending}.
\newblock P. Noordhoff, Ltd., Groningen, 1953.

\bibitem{22}
N.~I. Muskhelishvili.
\newblock {\em Singular integral equations. Boundary problems of function
  theory and their application to mathematical physics}.
\newblock Dover Publications, Inc., New York, 1992.

\bibitem{5}
B.M. Nuller.
\newblock On the deformation of an elastic wedge plate reinforced by a variable
  stiffness bar and a method of solving mixed problems.
\newblock {\em J. Appl. Math. Mech.}, 40(2):280--291, 1976.

\bibitem{2}
G.Y. Popov.
\newblock {\em Concentration of elastic stresses near stamps, cuts, thin
  inclusions and reinforcements}.
\newblock Nauka, Moscow, 1982.
\newblock in Russian.

\bibitem{13}
N.~Shavlakadze.
\newblock On singularities of contact stress upon tension and bending of plates
  with elastic inclusions.
\newblock {\em Proc. A. Razmadze Math. Inst.}, 120:135--147, 1999.

\bibitem{14}
N.~Shavlakadze.
\newblock The contact problem of bending of plate with thin fastener.
\newblock {\em Mechanic of solids}, 36:144–155, 2001.

\bibitem{15}
N.~Shavlakadze.
\newblock The contact problems of the mathematical theory of elasticity for
  plates with an elastic inclusion.
\newblock {\em Acta Appl. Math.}, 99(1):29--51, 2007.

\bibitem{16}
N.~Shavlakadze.
\newblock The solution of system of integral differential equations and its
  application in the theory of elasticity.
\newblock {\em ZAMM Z. Angew. Math. Mech.}, 91(12):979--992, 2011.

\bibitem{18}
N.~Shavlakadze.
\newblock Contact problem of electroelasticity for a piecewise homogeneous
  piezoelectric plate with an elastic coating.
\newblock {\em J. Appl. Math. Mech.}, 81(3):228--235, 2017.

\bibitem{17}
N.~Shavlakadze, Nana Odishelidze, and Francisco Criado-Aldeanueva.
\newblock The contact problem for a piecewise-homogeneous orthotropic plate
  with a finite inclusion of variable cross-section.
\newblock {\em Math. Mech. Solids}, 22(6):1326--1333, 2017.

\bibitem{10}
T.~C.~T. Ting.
\newblock Uniform stress inside an anisotropic elliptic inclusion with
  imperfect interface bonding.
\newblock {\em J. Elasticity}, 96(1):43--55, 2009.

\bibitem{11}
T.~C.~T. Ting and P.~Schiavone.
\newblock Uniform antiplane shear stress inside an anisotropic elastic
  inclusion of arbitrary shape with perfect or imperfect interface bonding.
\newblock {\em Internat. J. Engrg. Sci.}, 48(1):67--77, 2010.

\bibitem{9}
E.~Tsuchida, T.~Mura, and J.~Dundurs.
\newblock The elastic field of an elliptic inclusion with a slipping interface.
\newblock {\em Trans. ASME J. Appl. Mech.}, 53(1):103--107, 1986.

\end{thebibliography}

\end{document}